\documentclass[aip,reprint]{revtex4-1}
\usepackage{amsmath}
\usepackage{braket}
\usepackage{appendix}
\usepackage{graphicx}
\usepackage{dcolumn}
\usepackage{bm}
\usepackage[english]{babel}
\usepackage[mathlines]{lineno}
\usepackage[utf8]{inputenc}
\usepackage[T1]{fontenc}
\usepackage{mathptmx}
\usepackage{etoolbox}
\usepackage{xcolor}
\usepackage{nameref}
\usepackage{hyperref}
\draft 

\begin{document}


\title{Redox chemistry meets semiconductor defect physics}

\author{Jian Gu}
\affiliation{State Key Laboratory of Physical Chemistry of Solid Surfaces, iChEM, College of Chemistry and Chemical Engineering, Xiamen University, Xiamen 361005, China}

\author{Jun Huang}
\email[]{ju.huang@fz-juelich.de}
\affiliation{Institute of Energy and Technologies, IET-3: Theory and Computation of Energy Materials, Forschungszentrum Jülich GmbH, 52425 Jülich, Germany}
\affiliation{Theory of Electrocatalytic Interfaces, Faculty of Georesources and Materials Engineering, RWTH Aachen University, 52062 Aachen, Germany}

\author{Jun Cheng}
\email[]{chengjun@xmu.edu.cn}
\affiliation{State Key Laboratory of Physical Chemistry of Solid Surfaces, iChEM, College of Chemistry and Chemical Engineering, Xiamen University, Xiamen 361005, China}
\affiliation{Laboratory of AI for Electrochemistry (AI4EC), IKKEM, Xiamen 361005, China}
\affiliation{Institute of Artificial Intelligence, Xiamen University, Xiamen 361005, China}
\date{\today}
\begin{abstract}
Understanding how the electronic structure of electrodes influences electrocatalytic reactions has been a longstanding topic in the electrochemistry community, with predominant attention paid to metallic electrodes. In this work, we present a defect physics perspective on the effect of semiconductor band structure on electrochemical redox reactions. Specifically, the Haldane-Anderson model, originally developed to study multiple charge states of transition-metal defects in semiconductors, is extended to describe electrochemical redox reactions by incorporating the solvent effect, inspired by the Holstein model. 
The solvent coordinate and the actual charge on the redox species in reduced and oxidized states are assumed to be instant equilibrium, and the transitions between these states are defined by the framework of Green's function.
With these treatments, we treat the charge state transition in a self-consistent manner.
We first confirm that this self-consistent approach is essential to accurately depict the hybridization effect of band structure by comparing the model-calculated ionization potential (IP), electron affinity (EA), and redox potential of the species with those obtained from density functional theory (DFT) calculations. Next, we illustrate how this self-consistent treatment enhances our understanding of the catalytic activities of semiconductor electrodes and the source of asymmetry in reorganization energies, which is often observed in prior \textit{ab initio} molecular dynamics (AIMD) simulations. Additionally, we discuss how band structure impacts redox reactions in the strong coupling limit. Finally, we compare our work with other relevant studies in the literature.
\end{abstract}
\maketitle 

\section{Introduction}
For a free transition-metal atom in vacuum, the energy differences between different charge states are typically on the order of tens of eV.
The transitions between these charge states are difficult.
However, when the transition-metal atom is embedded in the semiconductor substrate, these energy differences can be significantly diminished to the order of 1 eV, enabling the defect to have multiple charge states then.
This phenomenon, observed in defect physics community, has primarily been attributed to the hybridization between the transition-metal defect and the band structure of semiconductor substrate\cite{haldane_simple_1976}.

In electrochemistry, if we view the aqueous species as a defect and the host solvent, mostly water, as an insulating oxide, the reduced state and oxidized state of the species can be considered as different charge states of the defect.
If we adopt the above defect physics perspective, the $-\text{IP}$ level (the ``$-$'' appears because the energy level is always negative, while the IP is always positive) of the reduced state and the $-\text{EA}$ level of the oxidized state can hybridize with the band structure of water in a similar way that the transition-metal defect does in a semiconductor substrate.
The redox potential can be approximated by averaging $-\text{IP}$ and $-\text{EA}$ in the Marcus picture\cite{marcus_electrostatic_1956,marcus_theory_1965,sato_electrochemistry_1998,memming_semiconductor_2015}, and will also be affected by hybridization.
This defect physics picture has been used to explain the source of errors in AIMD calculations of redox potentials\cite{adriaanse_aqueous_2012,cheng_redox_2014,liu_aqueous_2015,cheng_calculation_2016}.

The initial motivation of this work is to employ the above defect physics picture to revisit the catalytic effect of electrodes, which is currently understood largely through the channel of adsorption energy\cite{norskov_origin_2004}.
Here, we use a model Hamiltonian approach to study it.

Specifically, we build on the Haldane-Anderson model\cite{haldane_simple_1976}, which was originally proposed to describe the hybridization effect in defect physics. 
The model shows that hybridization allows for considerable variations in the oxidation state (formal charge) of the defect (\textit{i.e.}, the charge state can vary considerably), while the actual charge in the core regions of the transition-metal atom remains almost unchanged.
This hybridization effect was also interpreted as a ``charge self-regulation'' mechanism\cite{raebiger_charge_2008,dalpian_changes_2017}.
In addition, such a deviation of the actual charge from the formal charge induced by hybridization has also been studied by many researchers in defect physics\cite{quan_formal_2012,koch_density-based_2021}.

In this work, we will extend the Haldane-Anderson model\cite{haldane_simple_1976} to electrochemical systems by incorporating the solvent effect inspired by the Holstein model\cite{holstein_studies_1959}.
Specifically, a simple treatment proposed by Anderson\cite{anderson_model_1975} will be applied to the solvent potential, and a term similar to ``negative-\textit{U}''\cite{anderson_model_1975,coutinho_characterisation_2020} can be obtained in the Hamiltonian.
Combining this treatment with a compatible framework proposed by Haldane and Anderson\cite{haldane_simple_1976} of relating the actual charge to the oxidation state, the charge state transitions can be treated in a self-consistent manner.
Most importantly, we will show that this self-consistent treatment will lead to an effect on the actual charge of the aqueous species, which is very similar to the phenomenon of ``charge self-regulation'' in defect physics.
The consequent implications for electrocatalysis will be discussed.

The paper is organized as follows.
In Sec.~\ref{sec2a}, we introduce the basic aspects of the Haldane-Anderson model, which might not be well known to the electrochemistry community.
The incorporation of the solvent effect and the treatment of solvent potential proposed by Anderson\cite{anderson_model_1975} will be introduced in Sec.~\ref{sec2b}.
A compatible framework proposed by Haldane and Anderson\cite{haldane_simple_1976} to relate the actual charge and formal charge will be introduced in Sec.~\ref{sec2c}.
In this work, we will not calculate the potential energy surface of the redox reaction. 
Instead, we will regard the redox species as a defect, and calculate the $-\text{IP}$ level of its reduced state and $-\text{EA}$ level of its oxidized state.
To evaluate the hybridization effect on the thermodynamics of the redox reaction, the redox potential is needed.
In Sec.~\ref{sec2d}, we demonstrate the way to employ the thermodynamic integration (TI) method\cite{ferrario_redox_2006,sulpizi_acidity_2008,cheng_redox_2009} to compute it.
In Sec.~\ref{sec3}, we verify the applicability of our model in electrochemical systems by comparing the results computed by our model with that by DFT.
In Sec.~\ref{sec4a} through ~\ref{sec4c}, we discuss the insights gained from this defect physics picture.
In Sec.~\ref{sec4d}, we compare our work with related works in the literature.
We summarize our work in Sec.~\ref{sec5}.
\section{Model development}
\label{sec2}
\subsection{Elements of the Haldane-Anderson model}
\label{sec2a}
The Haldane-Anderson model was originally designed to address the problem of multiple charge states of transition-metal defects in semiconductors\cite{haldane_simple_1976}. 
Energy levels for different charge states of the defect after hybridization can be computed by the model, with the ``bare'' defect level as the input.
The model starts from the Anderson impurity Hamiltonian\cite{anderson_localized_1961}:
\begin{align}
    H&=E_d\sum\limits_{m\sigma}n_{m\sigma} 
    + \frac{1}{2}U\sum\limits_{\substack{m\sigma,\ m^{\prime}\sigma^{\prime}\\(m\sigma \ne m^{\prime}\sigma^{\prime})}} n_{m\sigma}n_{m^{\prime}\sigma^{\prime}} 
    + \sum\limits_{k\sigma} \epsilon_{k}n_{k\sigma}\nonumber\\
    &+ \sum\limits_{km\sigma}V_{mk}c_{m\sigma}^{\dagger}c_{k\sigma}
    +c.\ c.\ ,
    \label{eq1}
\end{align}
where $E_{d}$ is the ``bare'' defect level, $n_{m\sigma}$ is the occupancy operator of state $\ket{m\sigma}$, which is one of the 10 nearly degenerate $d$ orbitals for a transition-metal impurity. 
$U$ is the strong intra-atomic Coulomb repulsion between electrons localized in $d$ orbitals. 
$\epsilon_{k}$ is the dispersive energy of an electron in state $\ket{k\sigma}$ of the semiconductor substrate with an occupancy $n_{k\sigma}$. 
$V_{mk}$ is the interaction matrix element between extended states and the $d$ orbital of the defect.
$c_{m\sigma}^{\dagger}\ (c_{k\sigma})$ is the creation (annihilation) operator for the state $\ket{m\sigma}\ (\ket{k\sigma})$.
$c.\ c.$ in the last term is the abbreviation of complex conjugate. 

Under the Hartree-Fock approximation (HFA), a single-particle effective Hamiltonian can be obtained\cite{haldane_simple_1976,nolte_mesoscopic_1998,yamauchi_charge_2003}:
\begin{equation}
    H=E_{m\sigma}^{\text{eff}}n_{m\sigma}
    +\sum\limits_{k} \epsilon_{k} n_{k\sigma}
    +\sum\limits_{k} V_{mk} c_{m\sigma}^{\dagger} c_{k\sigma}
    +c.\ c.\ ,
    \label{eq2}
\end{equation}
where $E_{m\sigma}^{\text{eff}}$ is the effective defect level.
Here, one defect orbital $\ket{m\sigma}$ is considered, so the summation over orbitals is omitted.

The effective defect level $E_{m\sigma}^{\text{eff}}$ is a key parameter in the Hamiltonian and can be roughly understood as the defect level before considering the level broadening.
It should be determined self-consistently from
\begin{equation}
    E_{m\sigma}^{\text{eff}}=E_{d}+U\sum\limits_{\substack{m^{\prime}\sigma^{\prime}\\m^{\prime}\sigma^{\prime}\ne m\sigma}}\langle n_{m^{\prime}\sigma^{\prime}} \rangle.
    \label{eq3}
\end{equation}
The self-consistency requirement arises because when $E_{m\sigma}^{\text{eff}}$ of the defect changes relative to the Fermi level of the substrate, the charge of the defect $\sum\langle n_{m^{\prime}\sigma^{\prime}} \rangle$ will
change, then the potential felt by the electronic level will change due to Coulomb repulsion, which will in turn shift $E_{m\sigma}^{\text{eff}}$, and leading to a new charge.
In fact, the term $\langle n_{m\sigma}\rangle$ in Eq.~\ref{eq3} is the key to include the self-consistency effect.

The occupancy $\langle n_{m\sigma}\rangle$ in Eq.~\ref{eq3} can be self-consistently computed from Eq.~\ref{eq2} by
\begin{equation}
    \langle n_{m\sigma} \rangle=-\frac{1}{\pi}\text{Im}\int_{-\infty}^{\epsilon_{F}}G_{m\sigma}(\omega)\,d\omega.
    \label{eq4}
\end{equation}
Here, the $G_{m\sigma}(\omega)$ is the Green's function (See Appendix~\ref{apa} for the details about Green's function method)
\begin{equation}
    G_{m\sigma}(\omega)=\frac{1}{\omega -E_{m\sigma}^{\text{eff}}-\Sigma(\omega)}.
    \label{eq5}
\end{equation}

We determine $E_{m\sigma}^{\text{eff}}$ by solving Eqs.~\ref{eq3} to~\ref{eq5}.
In fact, the key to extending the Haldane-Anderson model to electrochemical systems lies in the modification of Eq.~\ref{eq3}, we will show this in Sec.~\ref{sec2b}.

\begin{figure}[htb]
    \centering
    \includegraphics{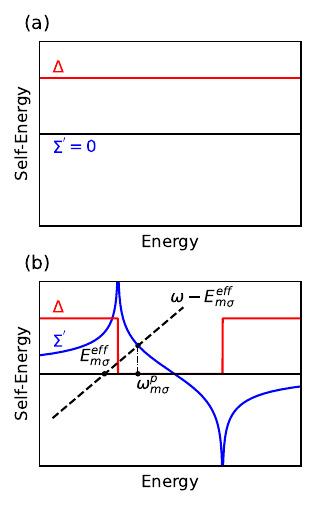}
    \caption{Schematic plot of the self-energy $\Sigma(\omega)$ of (a) metallic and (b) semiconductor substrate in WBA.}
    \label{fig1}
\end{figure}

The self-energy $\Sigma(\omega)$ in Eq.~\ref{eq5} is given by 
\begin{equation}
    \Sigma(\omega)=\sum\limits_{k} \frac{\lvert V_{km} \rvert ^{2}}{\omega -\epsilon_{k}},
    \label{eq6}
\end{equation}
which contains both real and imaginary parts:
\begin{equation}
    \Sigma(\omega)=\Sigma^{'}(\omega)-i\Delta(\omega).
    \label{eq7}
\end{equation}
For the metallic substrate, a constant density of states (DOS) and $k$, $m$ independent $V_{km}$ are always assumed, which is called the wide-band approximation (WBA). 
Under the WBA, the $\Delta(\omega)$ is constant according to
\begin{equation}
    \Delta(\omega)=\pi \lvert V_{km} \rvert ^{2} \rho_{k},
    \label{eq8}
\end{equation}
where $\rho_{k}$ is the DOS of the substrate (see Fig.~\ref{fig1}(a) for the shape of $\Sigma^{\prime}(\omega)$ and $\Delta(\omega)$ for metallic substrate in WBA).

\begin{figure}[htb]
    \centering
    \includegraphics[width=\linewidth]{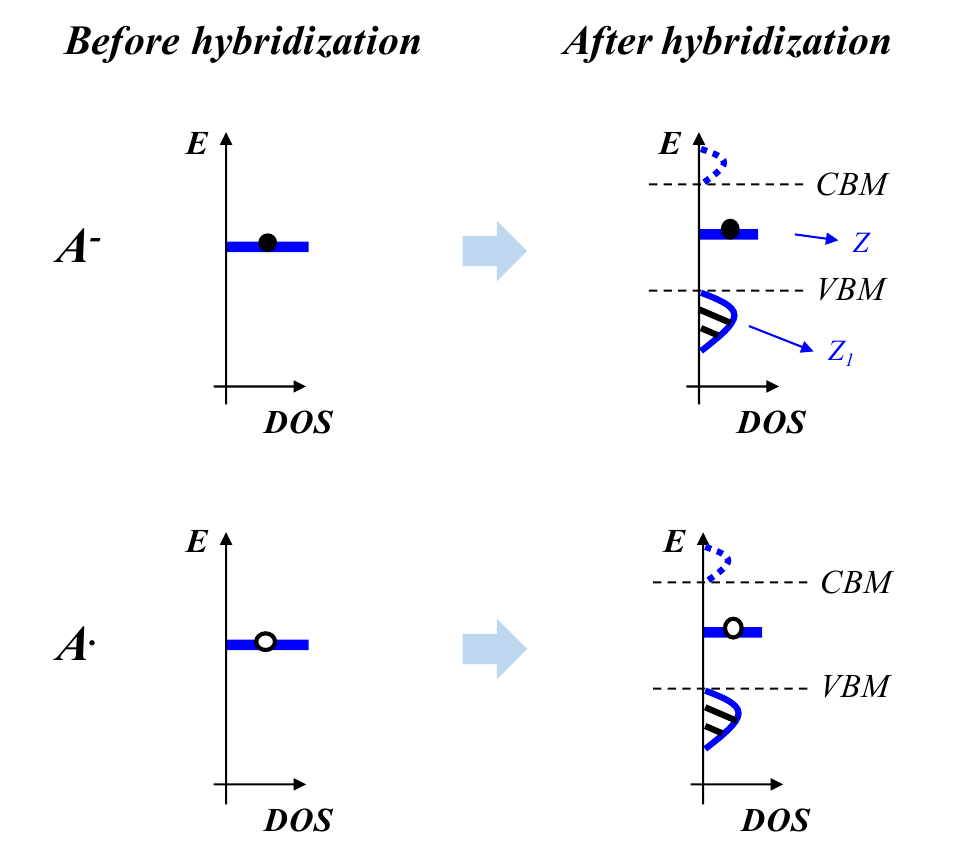}
    \caption{Schematic illustration of DOS projected on the defect site. 
    The figure also illustrates the approach for treating different charge states in this work. Before hybridization, the orbital (in blue) for $\text{A}^{-}$ and $\text{A}^{\bullet}$ are eigenstates of the system without broadening.
    The solid (hollow) dot on the level represents the state is occupied (unoccupied).
    After hybridization, the defect state has two parts: one is the bound state in the gap, and the other is the incoherent part due to hybridization with valence band (the part due to hybridization with conduction band is unoccupied and is denoted as dashed blue line).
    For both $\text{A}^{-}$ and $\text{A}^{\bullet}$, the incoherent part in the valence band are occupied; while for $\text{A}^{-}$, the bound state is occupied, and for $\text{A}^{\bullet}$, the bound state is unoccupied.}
    \label{fig2}
\end{figure}

For the semiconductor substrate, which is the focus of the Haldane-Anderson model and this work, the WBA is also used, where a gap will be introduced. 
Then, the real and imaginary parts of $\Sigma(\omega)$, are\cite{haldane_simple_1976}
\begin{equation}
    \Sigma^{'}(\omega)=\frac{\Delta}{\pi}\ln \lvert \frac{\omega-\epsilon_{c}}{\omega-\epsilon_{v}} \rvert
    \label{eq9}
\end{equation}
and
\begin{equation}
    \Delta(\omega)=\left\{
    \begin{array}{cc}
    \Delta, & \omega<\epsilon_{v}\ \text{or}\ \omega>\epsilon_{c},\\
    0, & \epsilon_{v}<\omega<\epsilon_{c},
    \end{array} \right.
    \label{eq10}
\end{equation}
respectively (see Fig.~\ref{fig1}(b) for the shape of $\Sigma^{'}(\omega)$ and $\Delta(\omega)$ for semiconductor substrate in WBA), where $\epsilon_{c}$ is the conduction band minimum (CBM) of the substrate, $\epsilon_{v}$ is the valence band maximum (VBM).
In this case, a pole of $G_{m\sigma} (\omega)$ in the gap will be formed (see Fig.~\ref{fig1}(b), the $\omega_{m\sigma}^{\text{p}}$ is the position of pole).
The resulting DOS projected on the defect can be referred to Fig.~\ref{fig2}. 
Before hybridization, the DOS is a $\delta$-function.
After hybridization, some part of the DOS remains the $\delta$-function in the gap.
Besides that, the DOS also has parts due to hybridization with conduction band and valence band.
This suggests that the occupancy of the defect orbital has two contributions, one is the residue of the pole in the gap, the other is the incoherent part due to hybridization with valence band (the part due to hybridization with conduction band is unoccupied, so has no contribution to the occupancy). 
The residue of the pole (see Sec.~\ref{sec2c}) gives the weight of the eigenstate resembling that before hybridization\cite{khomskii_basic_2010}, which is a bound state localized in the defect site.
The incoherent part stands for the charge localized in the long hydrogenic tail regions of the defect wave function\cite{haldane_simple_1976}.
The incoherent part and the bound state of the defect orbital can be roughly understood as the bonding and antibonding orbitals, respectively, in a molecular orbital picture.
Both of them belong to the defect, but the shape can be different.

Hereafter, the charge of the defect, which contains both the bound state charge and the incoherent part, will be called more precisely the \textit{actual charge}.
It is a sum of the occupancy of all the orbitals, which can be computed by Eq.~\ref{eq4}.
The actual charge determines how the added electron will be repelled, or how the solvents will respond, see below.
We point out that the actual charge has to be defined by some projection approaches, \textit{e.g.}, projection to the defect site as is done in the Haldane-Anderson model (see Sec.~\ref{sec2c}), or the commonly used Mulliken population analysis\cite{mulliken_electronic_1955} in \textit{ab initio} calculations.
\subsection{Incorporation of solvent effect}
\label{sec2b}
\begin{figure}[htb]
    \centering
    \includegraphics[width=\linewidth]{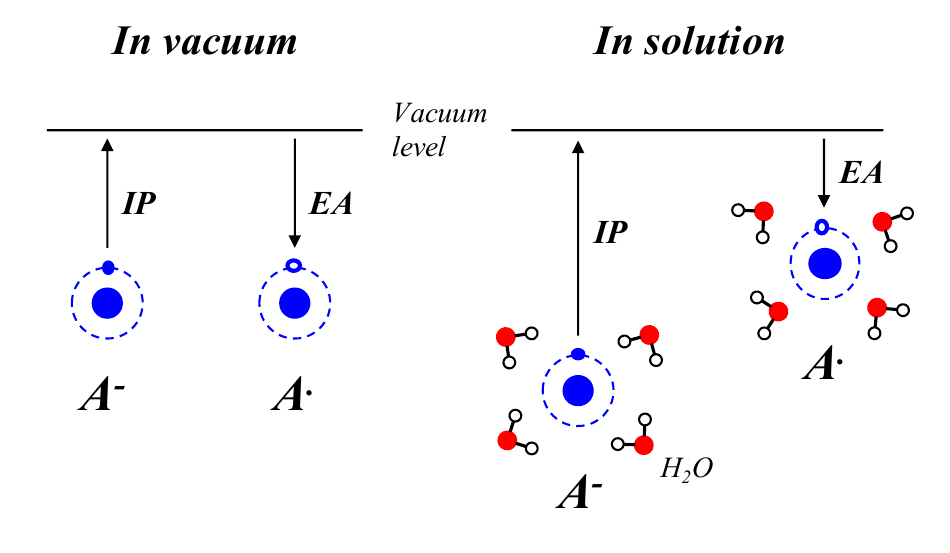}
    \caption{Schematic illustration of the relationship between HOMO of $\text{A}^{-}$ and LUMO of $\text{A}^{\bullet}$, where A is the aqueous species like $\text{OH}^{-}$, $\text{Cl}^{-}$, etc. 
    The physical meaning of IP and EA are also shown schematically in the figure. 
    All the orbitals are denoted as blue dashed circle.}
    \label{fig3}
\end{figure}

As regards electrochemical redox reactions, the solvent effect must be considered in computing the energy levels.
According to the Marcus-Gerischer model\cite{sato_electrochemistry_1998,memming_semiconductor_2015}, the adjustment of solvent polarization according to the \textit{actual charge} of the species splits the $-\text{IP}$ level of of $\text{A}^{-}$ from $-\text{EA}$ level of $\text{A}^{\bullet}$ (see Fig.~\ref{fig3}, here A denotes the aqueous species).
In fact, the $-\text{IP}$ level of $\text{A}^{-}$ is the highest occupied molecular orbital (HOMO) according to Koopmans’ theorem\cite{luo_koopmans_2006}, and the $-\text{EA}$ level of $\text{A}^{\bullet}$ can also be approximated by the lowest unoccupied molecular orbital (LUMO).
In vacuum, the HOMO of $\text{A}^{-}$ and LUMO of $\text{A}^{\bullet}$ is the same orbital, since if we remove the electron on HOMO of $\text{A}^{-}$, the species will become $\text{A}^{\bullet}$, and the orbital is now the LUMO of the new species $\text{A}^{\bullet}$ (see Fig.~\ref{fig3}).

We first see how the solvent effect is treated in the literature.
In Schmickler's theoretical model for electron transfer reactions on both metallic and semiconductor electrodes\cite{schmickler_theory_1986,santos_theory_2012,schmickler_adiabatic_2017}, the Hamiltonian is:
\begin{equation}
    H=H_{\text{el}}+\lambda q^{2}+(1-n_{m\sigma})2\lambda q,
    \label{eq11}
\end{equation}
where $H_{\text{el}}$ is the Anderson impurity Hamiltonian Eq.~\ref{eq2} as used in this work, $\lambda$ is the solvent reorganization energy, $q$ is the solvent coordinate, and $n_{m\sigma}$ is the occupancy operator of the aqueous species.
The solvent effect is taken into account by the charge-solvent coupling term $-2\lambda n_{m\sigma}q$ and the solvent potential $\lambda q^{2}+2\lambda q$.
This treatment of charge-solvent coupling is similar to that of the electron-lattice coupling in the Holstein model\cite{holstein_studies_1959,li_ground-state_2010,franchini_polarons_2021}, which is usually used to describe small polarons in condensed matter physics.
The Holstein model is:
\begin{equation}
    H=-t\sum_{i,j}(c_{i}^{\dagger}c_{j}+c_{j}^{\dagger}c_{i})+\sum_{i}(\frac{p_{i}^{2}}{2M}+\frac{1}{2}M\omega_{E}^{2}x_{i}^{2})-\alpha\sum_{i}n_{i}x_{i},
    \label{eq12}
\end{equation}
where the first term describes electron hopping, $\frac{p_{i}^{2}}{2M}$ describes the kinetic energy of the ion, $\frac{1}{2}M\omega_{E}^{2}x_{i}^{2}$ describe the ion potential, and the last term describes the electron-lattice coupling, in which $x_{i}$ is the ion displacement, and $\alpha$ is the coupling constant.
In fact, it has already been pointed out that the Marcus theory and the Holstein model are equivalent for small polarons\cite{deskins_electron_2007,wang_constrained_2021}.

Similar to Schmickler's model, here we also use the Holstein model to describe the solvent effect.
But we will show that, by a treatment proposed by Anderson\cite{anderson_model_1975}, a $\langle n_{m\sigma}\rangle$ term can appear in the effective defect level $E_{m\sigma}^{\text{eff}}$, and the model can incorporate the self-consistency effect even without considering the Coulomb repulsion term.

In the Holstein model\cite{holstein_studies_1959,li_ground-state_2010}, the potential energy exerted by the solvent is:
\begin{equation}
    V=\frac{1}{2}M\omega_{E}^{2}x_{i}^{2}-\alpha x_{i}n_{i}.
    \label{eq13}
\end{equation}
In the vicinity of the equilibrium state, $\frac{\partial V}{\partial x_{i}}=0$, this leads to
\begin{equation}
    x_{i}=\frac{\alpha}{M\omega_{E}^{2}}n_{i},
    \label{eq14}
\end{equation}
substituting into Eq.~\ref{eq13}, we have
\begin{equation}
    V=-\frac{\alpha^{2}}{2M\omega_{E}^{2}}n_{i}-\frac{\alpha^{2}}{M\omega_{E}^{2}}n_{i\uparrow}n_{i\downarrow}.
    \label{eq15}
\end{equation}
We see that an electron-electron interaction term appears with this simple treatment, which is similar to the Coulomb repulsion term.
In fact, the above treatment follows the idea of Anderson\cite{anderson_model_1975}.
With this observation, he introduced the concept of ``negative-\textit{U}''\cite{anderson_model_1975,coutinho_characterisation_2020}.
Now, the full Hamiltonian containing the \textit{equilibrium} solvent effect can be written as
\begin{align}
    H&=H_{\text{el}}+V\nonumber\\
    &=(E_{d}-\frac{\alpha^{2}}{2M\omega_{E}^{2}})\sum_{m\sigma}n_{m\sigma}-\frac{\alpha^{2}}{M\omega_{E}^{2}}\sum_{i}n_{i\uparrow}n_{i\downarrow}+\sum_{k\sigma}\epsilon_{k}n_{k\sigma}\nonumber\\
    &+\sum_{km\sigma}V_{mk}c_{m\sigma}^{\dagger}c_{k\sigma}+c.\ c.
\end{align}
Here, the Coulomb repulsion term is dropped, since we mainly focus on $2p$ and $3p$ orbitals in electrochemistry.
Under the HFA, we can also have the single-particle effective Hamiltonian of the form of Eq.~\ref{eq2}, but with a different $E_{m\sigma}^{\text{eff}}$:
\begin{equation}
    E_{m\sigma}^{\text{eff}}=E_{d}-\frac{\alpha^{2}}{2M\omega_{E}^{2}}-\frac{\alpha^{2}}{M\omega_{E}^{2}}\langle n_{m\sigma^{\prime}}\rangle.
\end{equation}
Here, we only consider one of the four pairs of $sp^{3}$ electrons in closed shell $\text{OH}^{-}$, $\text{Cl}^{-}$, etc. 
If we set $\frac{\alpha^{2}}{2M\omega_{E}^{2}}=\lambda$, and $E_{d}-3\lambda=-\text{IP}_{\text{unhyb}}$, we have
\begin{equation}
    E_{m\sigma}^{\text{eff}}=-\text{IP}_{\text{unhyb}}+2\lambda(1-\langle n_{m\sigma^{\prime}}\rangle).
\end{equation}
Here, we recognize that $-\text{IP}_{\text{unhyb}}$ has the meaning of $-\text{IP}$ level of the defect before hybridization.
To make things easier, we assume $E_{m\sigma}=E_{m\sigma^{\prime}}$, \textit{i.e.}, a spin unpolarized case, we finally have
\begin{equation}
    E_{m\sigma}^{\text{eff}}=-\text{IP}_{\text{unhyb}}+2\lambda(1-\langle n_{m\sigma}\rangle).
    \label{eq19}
\end{equation}
Now, we can focus on just one spin-orbital of the species.

We can see that with above treatment, the occupancy of the orbital $\langle n_{m\sigma}\rangle$ determines the effective defect level $E_{m\sigma}^{\text{eff}}$ on one hand, and it has to be determined by solving the Hamiltonian on the other hand.
This is similar to the case in defect physics Eq.~\ref{eq3}, requiring a self-consistent treatment of the occupancy of the orbital.
In fact, in electrochemistry, the self-consistency requirement also arises naturally: when $E_{m\sigma}^{\text{eff}}$ of the aqueous species changes relative to the Fermi level of the electrode, the actual charge of the species will
change, then the solvent will respond to this new charge, which will in turn shift $E_{m\sigma}^{\text{eff}}$, and leading to a new charge.

We emphasize that Eq.~\ref{eq19} describes the reduced and oxidized states with their respective equilibrium solvation. The transition between them will be achieved by the technique of TI, see Sec.~\ref{sec2d}.

Here we make a further approximation.
When the defect is adsorbing on the surface of electrodes, the reorganization of the environment will include not only the solvation shell but also the electrode material\cite{cheng_hole_2012,cheng_identifying_2014,cheng_reductive_2015}.
So the reorganization energy $\lambda$ at the interface will be different from that in the bulk water\cite{huang_mixed_2020}.
To make things easier, here, we will use the value of $\lambda$ for bulk water as an approximation for the value at the interface. 
It's noteworthy that for many transition-metal oxide electrodes of interest, their dielectric constants are on the order of that of bulk water (e.g., $\sim 100$ for $\text{TiO}_{2}$\cite{parker_static_1961}, $\sim 20$ for $\text{Fe}_{2}\text{O}_{3}$\cite{lunt_dielectric_2013}, compared to $\sim 78$ for bulk water).
So, the above approximation is not expected to alter the main physical insights we aim to convey in this work, although it may introduce some numerical discrepancies.
\subsection{Assigning the actual charge to reduced and oxidized states}
\label{sec2c}
According to Eq.~\ref{eq19}, to compute the energy levels for oxidized/reduced state, the corresponding occupancy of the orbital $\langle n_{m\sigma}\rangle$ (or actual charge, since we only consider one spin-orbital of the species) is needed.
However, the reduced and oxidized states are defined by the oxidation state (formal charge).
How to assign the actual charge to reduced and oxidized states?

In many cases, the formal charge is a good approximation for the actual charge (e.g., for an isolated $\text{Fe}^{3+}$ ion, its actual charge is +3 electronic charges, and its formal charge is also +3).
However, when the defect hybridizes with materials with extended states, the relationship between the actual charge and the formal charge is nontrivial.

Here, we will use a framework to relate the actual charge to the formal charge, which is compatible with our self-consistent treatment of actual charge.
This framework follows the idea of Haldane and Anderson\cite{haldane_simple_1976}.
The basic idea is: hybridization with materials with extended states will transfer some atomic charge of the defect into incoherent part, and the rest will remain bound (see Fig.~\ref{fig2}).
In this case, when we attempt to add 1 electron to the defect, the electron will be partially delocalized due to hybridization, and what we really added is the bound state charge.
At the same time, the incoherent part will also contribute to the actual charge, but in a self-consistent way, since the increased defect charge will raise the electronic level due to the Coulomb repulsion term (or lower the level due to the solvent reorganization), which in turn results in decrease of the actual charge.

Therefore, the change in formal charge in fact corresponds to the change in the \textit{number of occupied bound states}.
With this consideration, the actual charge of $\text{A}^{-}$ can be computed by Green's function
\begin{equation}
    \langle n_{m\sigma} \rangle_{\text{A}^{-}}=Z_{1}(E_{m\sigma}^{\text{eff}})+Z(E_{m\sigma}^{\text{eff}}),
    \label{eq20}
\end{equation}
where
\begin{equation}
    Z_{1}(E_{m\sigma}^{\text{eff}})=-\frac{1}{\pi}\text{Im}\int_{-\infty}^{\epsilon_{v}}G_{m\sigma}(\omega)\,d\omega,
    \label{eq21}
\end{equation}
\begin{equation}
    Z(E_{m\sigma}^{\text{eff}})=-\frac{1}{\pi}\text{Im}\int_{\epsilon_{v}}^{\epsilon_{c}}G_{m\sigma}(\omega)\,d\omega.
    \label{eq22}
\end{equation}
Here, $Z_{1}(E_{m\sigma}^{\text{eff}})$ is the contribution from the incoherent part, and $Z(E_{m\sigma}^{\text{eff}})$ is the residue of the pole of $G_{m\sigma}(\omega)$ (bound state charge).
Substituting Eqs.~\ref{eq9} and~\ref{eq10} into Eq.~\ref{eq21}, we have
\begin{equation}
    Z_{1}(E_{m\sigma}^{\text{eff}})=\frac{1}{\pi}\int_{-\infty}^{\epsilon_{v}}\frac{\Delta}{[\omega-E_{m\sigma}^{\text{eff}}-\Sigma^{'}(\omega)]^{2}+\Delta^{2}}\,d\omega,
    \label{eq23}
\end{equation}
and $Z(E_{m\sigma}^{\text{eff}})$ can calculated by\cite{haldane_simple_1976}
\begin{equation}
    Z(E_{m\sigma}^{\text{eff}})=\frac{1}{\lvert 1-\frac{d}{d\omega}\Sigma^{'}(\omega) \rvert}_{\omega_{m\sigma}^{\text{p}}}.
    \label{eq24}
\end{equation}

Similarly, for $\text{A}^{\bullet}$, the bound state is now unoccupied, but there is still an incoherent part of $\ket{m\sigma}$.
So the actual charge of $\text{A}^{\bullet}$ is
\begin{equation}
    \langle n_{m\sigma} \rangle_{\text{A}^{\bullet}}=Z_{1}(E_{m\sigma}^{\text{eff}}).
    \label{eq25}
\end{equation}

Now, we can substitute Eq.~\ref{eq20} and Eq.~\ref{eq25} into Eq.~\ref{eq19} for $E_{m\sigma}^{\text{eff}}$ of reduced state $\text{A}^{-}$ and oxidized state $\text{A}^{\bullet}$:
\begin{equation}
    E_{\text{A}^{-}}^{\text{eff}}=-\text{IP}_{\text{unhyb}}+2\lambda(1-Z_{1}(E_{\text{A}^{-}}^{\text{eff}})-Z(E_{\text{A}^{-}}^{\text{eff}})),
    \label{eq26}
\end{equation}
\begin{equation}
    E_{\text{A}^{\bullet}}^{\text{eff}}=-\text{IP}_{\text{unhyb}}+2\lambda(1-Z_{1}(E_{\text{A}^{\bullet}}^{\text{eff}})).
    \label{eq27}
\end{equation}

By this point, we have extended the Haldane-Anderson model to electrochemical systems.
Substituting Eqs.~\ref{eq26} and~\ref{eq27} into Eq.~\ref{eq2}, we obtain the final Hamiltonian.
To solve the Hamiltonian, the effective defect levels Eqs.~\ref{eq26} and~\ref{eq27} should be computed self-consistently firstly.
Since we use the HOMO of $\text{A}^{-}$ to approximate $-\text{IP}$ and the LUMO of $\text{A}^{\bullet}$ for $-\text{EA}$, the DOS of the orbital is needed.
We will use Green's function method to compute the DOS (see Appendix~\ref{apa}), and then we can compute the $-\text{IP}$ and $-\text{EA}$.
\subsection{Calculation of redox potential by thermodynamic integration}
\label{sec2d}
To understand how the hybridization affects the thermodynamics of redox reactions, we need to calculate the free energy change (redox potential $U^{o}$) of the redox reaction.
The $U^{o}$ is an adiabatic level, which is not equal to $-\text{IP}$ or $-\text{EA}$, both of which are vertical levels.
In fact, with $-\text{IP}$ and $-\text{EA}$, the $U^{o}$ can be approximated by $U^{o}=1/2(-\text{IP}-\text{EA})$, according to the Marcus-Gerischer model\cite{sato_electrochemistry_1998,memming_semiconductor_2015}.
However, this presupposes that $\lambda_{O}^{\prime}=\lambda_{R}^{\prime}$ even after hybridization (here, $\lambda_{O}$ and $\lambda_{R}$ are reorganization energies for oxidation and reduction processes, respectively, and the prime $\lambda^{\prime}$ denotes that the reorganization energy is the computed one, to be distinguished from the one as model input $\lambda$).
However, we will show that even we assume the solvent responds linearly with the charge (we use this assumption in our model since we start from the Holstein model), the resulting $\lambda_{O}^{\prime}$ may not be equal to $\lambda_{R}^{\prime}$ due to hybridization.
To demonstrate this, we should calculate the $U^{o}$ more precisely by TI method\cite{ferrario_redox_2006,sulpizi_acidity_2008,cheng_redox_2009}.

In TI method, to calculate the free energy change during the transformation from state $i$ to state $j$, a linear coupling Hamiltonian is always constructed:
\begin{equation}
    H_{\eta}=\eta H_{j}+(1-\eta)H_{i},
    \label{eq28}
\end{equation}
where $H_{j}$ and $H_{i}$ are the corresponding Hamiltonian for state $j$ and $i$, respectively.
By varying the coefficient $\eta$ from 0 to 1, we creates a series of potential energy surfaces transforming from $H_{i}$ to $H_{j}$ gradually.

The link of this linear coupling Hamiltonian with the free energy change can be found from the definition of free energy:
\begin{equation}
    F(\eta)=-k_{B}T \ln \Lambda ^{-3N} \int \exp (-\beta E_{\eta})\,dR^{N},
    \label{eq29}
\end{equation}
where $E_{\eta}$ is the potential energy surface under the Hamiltonian $H_{\eta}$, with atomic position
$R^{N}$.
$\Lambda$ is the average thermal wavelength of the atoms, $\beta^{-1}=k_{B}T$ with $k_{B}$ the Boltzmann constant and $T$ the temperature.
From Eq.~\ref{eq29}, the free energy change is
\begin{align}
    \Delta F(\eta)&=F(\eta+\Delta \eta)-F(\eta)\nonumber\\
    &=-k_{B}T\ln \langle \exp [-\beta (E_{\eta +\Delta \eta}-E_{\eta})] \rangle_{\eta},
    \label{eq30}
\end{align}
where $\langle\ldots\rangle$ is the average over the potential energy surface of $H_{\eta}$.
According to Eq.~\ref{eq28}, we notice that
\begin{align}
    E_{\eta +\Delta \eta}-E_{\eta}&=\{(\eta+ \Delta \eta)E_{j}+[1-(\eta+ \Delta\eta)]E_{i}\}\nonumber\\
    &-[\eta E_{j}+(1-\eta)E_{i}]\nonumber\\
    &=\Delta E \Delta \eta,
    \label{eq31}
\end{align}
where $\Delta E=E_{j}-E_{i}$ is the vertical energy gap with the same atomic position $R^{N}$.
This suggests that Eq.~\ref{eq30} can be rewritten as
\begin{equation}
    dF(\eta)=\langle\Delta E\rangle_{\eta} d\eta.
    \label{eq32}
\end{equation}
Now we can obtain the familiar expression for TI:
\begin{align}
    \Delta F&=F(1)-F(0)\nonumber\\
    &=\int_{0}^{1}\frac{dF(\eta)}{d\eta}\,d\eta\nonumber\\
    &=\int_{0}^{1}\langle\Delta E\rangle_{\eta}\,d\eta.
    \label{eq33}
\end{align}
This is the final equation used to calculate free energy change in TI method.
The vertical energy gap $\langle \Delta E \rangle_{\eta}$ should be computed under the linear coupling Hamiltonian of Eq.\ref{eq28}.

To implement the TI method in our model, the key is the construction of the linear coupling Hamiltonian.
In this work, the $i$ state is the $\text{A}^{-}$ state, and $j$ state is the $\text{A}^{\bullet}$ state.
By Eq.~\ref{eq28}, what we really constructed is a fictitious defect, whose interaction with solvents is linearly coupled to that of $\text{A}^{-}$ and $\text{A}^{\bullet}$.
In the potential energy surface of $\text{A}^{-}$, the defect tends to reorganize the solvent with the energy of $2\lambda (1-Z_{1}-Z)$ (see Eq.~\ref{eq26}); while in the potential energy surface of $\text{A}^{\bullet}$, the defect tends to reorganize the solvent with the energy of $2\lambda (1-Z_{1})$ (see Eq.~\ref{eq27}).
This suggests that the linear coupling Hamiltonian can be constructed as:
\begin{align}
    H_{\eta}&=\{-\text{IP}_{\text{unhyb}}+2\lambda [1-Z_{1}-Z(1-\eta)]\}n_{m\sigma}\nonumber\\
    &+\sum\limits_{k}\epsilon_{k}n_{k\sigma}+\sum\limits_{k}V_{km}c_{k\sigma}^{\dagger}c_{m\sigma}+\text{c. c.}
    \label{eq34}
\end{align}

After the construction of the Hamiltonian, the procedure in Appendix~\ref{apa} is again used to compute the energy levels. 
Then, the vertical energy gap $\langle \Delta E \rangle_{\eta}$ can be calculated, and the $\Delta F$ and $U^{o}$ can be obtained.
\section{Results}
\label{sec3}
\subsection{Hybridization with the band structure of titanium dioxide}
\label{sec3a}
We first consider the redox reaction
\begin{equation}
    \text{OH}^{-}\rightarrow \text{OH}^{\bullet}+\text{e}^{-}
    \label{eq35}
\end{equation}
hybridized with the band structure of $\text{TiO}_{2}$.
As introduced in Sec.~\ref{sec2}, we use $-\text{IP}$ of $\text{OH}^{-}$ before hybridization, \textit{i.e.}, the $-\text{IP}$ of $\text{OH}^{-}$ in bulk water, and the corresponding reorganization energy $\lambda$, as model inputs.
Both of them can be obtained from experiments.
To describe the band structure of $\text{TiO}_{2}$, the WBA will be used, with VBM and CBM also obtained from experiments.
All the model inputs used in this work can be found in Tables~\ref{tab1} and ~\ref{tab2}.
\begin{table}[htb]
\caption{\label{tab1}Model inputs for IP of $\text{A}^{-}$ and $\lambda$ before hybridization. All the units are eV.}
\begin{ruledtabular}
\begin{tabular}{cccc}
& Cl & OH & $\text{HO}_{2}$ \\
\hline
IP & 9.6\footnotemark[1] & 9.2\footnotemark[1] & 7.1\footnotemark[2] \\
$\lambda$ & 2.9\footnotemark[3] & 2.9\footnotemark[3] & 2.4\footnotemark[2] \\
\end{tabular}   
\end{ruledtabular}
\footnotetext[1]{Experimental values, from Ref.~\onlinecite{winter_electron_2006}.}
\footnotetext[2]{RPA values, from Ref.~\onlinecite{cheng_calculation_2016}.}
\footnotetext[3]{Estimated by $\lambda=\lvert\text{IP}\rvert-\lvert U^{o}\rvert$, where $U^{o}$ are experimental values from Ref.~\onlinecite{stanbury_reduction_1989}.}
\end{table}
\begin{table}[htb]
\caption{\label{tab2}Model inputs for CBM and VBM. All the values are referenced to vacuum in V.}
\begin{ruledtabular}
\begin{tabular}{ccc}
& CBM & VBM \\
\hline
$\text{TiO}_{2}$ (exp.) & -4.1\footnotemark[1] & -7.1\footnotemark[1] \\
Bulk water (at PBE0) & -1.2\footnotemark[2] & -8.2\footnotemark[2] \\
Bulk water (at HSE06) & -1.4\footnotemark[2] & -8.1\footnotemark[2] \\
Bulk water (at BLYP) & -1.9\footnotemark[2] & -6.8\footnotemark[2] \\
\end{tabular}
\end{ruledtabular}
\footnotetext[1]{Experimental values, from Refs.~\onlinecite{gratzel_photoelectrochemical_2001,cheng_aligning_2010}.}
\footnotetext[2]{From Ref.~\onlinecite{cheng_calculation_2016}.}
\end{table}

\begin{figure}[htb]
    \centering
    \includegraphics{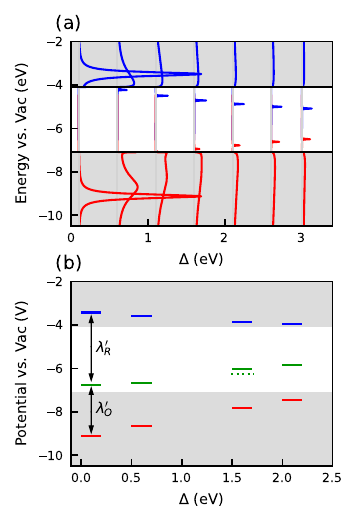}
    \caption{(a) DOS of HOMO of $\text{OH}^{-}$ (in red color) and LUMO of $\text{OH}^{\bullet}$ (in blue color) as a function of $\Delta$. (b) Level diagram of OH on $\text{TiO}_{2}$ as a function of $\Delta$.
    The definition of $\Delta$ can be found in Eq.~\ref{eq8}; it represents the electronic coupling between the redox couple and the substrate.
    The solid red, green, and blue lines correspond to the $-\text{IP}$ level of $\text{OH}^{-}$, $U^{o}$ of the couple, and $-\text{EA}$ level of $\text{OH}^{\bullet}$, respectively.
    The prime on $\lambda$ denotes that it is a computed reorganization energy, to be distinguished from the one as model inputs.}
    \label{fig4}
\end{figure}

Fig.~\ref{fig4}(a)
shows the calculated DOS of HOMO of $\text{OH}^{-}$ (red color) and LUMO of $\text{OH}^{\bullet}$ (blue color) after hybridization with $\text{TiO}_{2}$'s band structure.
Here, we treat the electronic coupling between the redox ion and the substrate, $\Delta$, as a parameter (see Eq.~\ref{eq8} for the definition of $\Delta$), as in the work of Haldane and Anderson\cite{haldane_simple_1976}.
It can be seen from Fig.~\ref{fig4}(a) that as $\Delta$ increases, a bound state emerges within the gap for both HOMO of $\text{OH}^{-}$ and LUMO of $\text{OH}^{\bullet}$.
This phenomenon is consistent with DFT calculations, which also reveal a similar gap state when $\text{OH}^{\bullet}$ adsorbs on the surface of $\text{TiO}_{2}$\cite{di_valentin_bulk_2011,cheng_identifying_2014}.
In those DFT studies, this gap state was considered to arise from the localization of the hole at $\text{OH}^{\bullet}$ in the form of a small polaron.
Here, the origin of such gap states can be elucidated more clearly by our model Hamiltonian approach. 
We find that the formation of small polarons is not the sole factor responsible for these gap states; hybridization with the semiconductor’s band structure also plays a significant role (see Fig.~\ref{fig1}(b)).
Besides, we notice that the bound state of LUMO of $\text{OH}^{\bullet}$ is more pronounced than that of HOMO of $\text{OH}^{-}$.
The origin of this difference is apparent in our model, which can be attributed to the smaller energy separation between the level of LUMO of $\text{OH}^{\bullet}$ and CBM, compared to that between HOMO of $\text{OH}^{-}$ and VBM.

According to DFT calculations, the level of the bound state of LUMO of $\text{OH}^{\bullet}$ is 2.4 eV above the VBM of $\text{TiO}_{2}$\cite{cheng_identifying_2014}.
Using this value, we can estimate the $\Delta$ in our model to be 1.6 eV (see Fig.~\ref{fig4}(a), all the fitted values of $\Delta$ in this work can be found in Table~\ref{tab3}, the computed actual charge can be found in Table~\ref{tab4}).
\begin{table}[htb]
\caption{\label{tab3}Fitted values of $\Delta$ for various redox species (Cl, OH, and $\text{HO}_{2}$) interact with various substrates ($\text{TiO}_{2}$, with band edges from experiments; and bulk water, with band edges calculated at different functionals).
The units are eV.}
\begin{ruledtabular}
    \begin{tabular}{cccc}
    & Cl & OH & $\text{HO}_{2}$ \\
    \hline
$\text{TiO}_{2}$ & - & 1.6 & - \\
Bulk water (at PBE0) & 1.2 & 2 & 0.4 \\
Bulk water (at HSE06) & 2 & 2.8 & 0.5 \\
Bulk water (at BLYP) & 0.5 & 0.4 & 2 \\
    \end{tabular}
\end{ruledtabular}
\end{table}
\begin{table}[htb]
\caption{\label{tab4}Computed actual charge (Eq.~\ref{eq4}) for various redox species (Cl, OH, and $\text{HO}_{2}$) interact with various substrates ($\text{TiO}_{2}$, with band edges from experiments; and bulk water, with band edges calculated at different functionals).}
\begin{ruledtabular}
    \begin{tabular}{ccccccc}
    & $\text{Cl}^{-}$ & $\text{Cl}^{\bullet}$ & $\text{OH}^{-}$ & $\text{OH}^{\bullet}$ & $\text{HO}_{2}^{-}$ & $\text{HO}_{2}^{\bullet}$ \\
    \hline
$\text{TiO}_{2}$ & - & - & 0.87 & 0.11 & - & - \\
Bulk water (at PBE0) & 0.93 & 0.06 & 0.89 & 0.09 & 0.97 & 0.02 \\
Bulk water (at HSE06) & 0.89 & 0.10 & 0.84 & 0.11 & 0.96 & 0.02 \\
Bulk water (at BLYP) & 0.97 & 0.04 & 0.97 & 0.03 & 0.85 & 0.10 \\
    \end{tabular}
\end{ruledtabular}
\end{table}
With this $\Delta$, the redox potential $U^{o}$ predicted by our model is $-6.0$ eV (solid green line in Fig.~\ref{fig4}(b)).
Remarkably, this value is very close to that obtained from DFT calculations ($-6.25$ eV\cite{cheng_identifying_2014}, dashed green line in Fig.~\ref{fig4}(b)).
Such close agreement between the $U^{o}$ calculated by our model and that by DFT strongly suggests that our model can effectively capture the hybridization effect of the semiconductor band structure on redox reactions. 
To further solidify this conclusion, in next section, we will investigate redox reactions hybridized with the band structure of bulk water and also compare our results with DFT.
\subsection{Hybridization with the band structure of bulk water}
\label{sec3b}
\begin{figure*}[htb]
    \centering
    \includegraphics{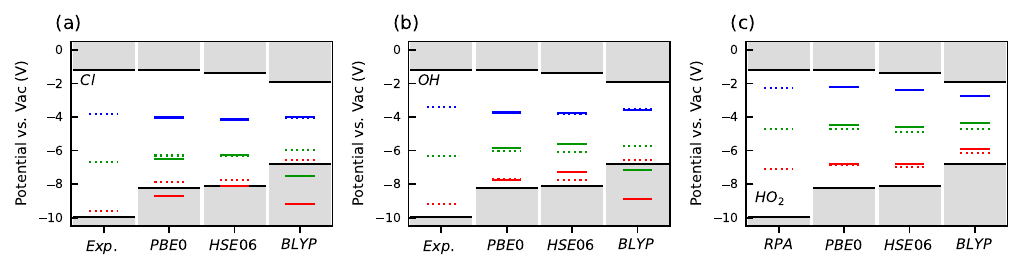}
    \caption{Level diagram of redox ion A interacts with the band structure of bulk water at different functionals, with A denotes (a) Cl, (b) OH and (c) $\text{HO}_{2}$.
    The solid red, green, and blue lines correspond to the $-\text{IP}$ level of $\text{A}^{-}$, $U^{o}$ of the couple $\text{A}^{-}/\text{A}^{\bullet}$, and $-\text{EA}$ level of $\text{A}^{\bullet}$ calculated by our model, respectively.
    The dashed lines correspond to the values taken from DFT (the data is taken from Refs.\onlinecite{cheng_calculation_2016,cheng_redox_2014}).
    The band structure of bulk water are calculated by PBE0, HSE06 and BLYP as denoted in the figure, with corresponding VBM and CBM plotted by black solid lines.}
    \label{fig5}
\end{figure*}

Here, we calculate the $-\text{IP}$ of $\text{A}^{-}$, $-\text{EA}$ of $\text{A}^{\bullet}$ and $U^{o}$ of $\text{A}^{-}/\text{A}^{\bullet}$ after hybridization with bulk water, where A represents OH, Cl and $\text{HO}_{2}$.
The experimental values of $-\text{IP}$ of $\text{A}^{-}$ in bulk water serve as the ``bare'' defect level (this choice is justified in next paragraph).
When the experimental values of $-\text{IP}$ and $\lambda$ are not available, the values calculated by random phase approximation (RPA) will be used instead as model inputs, since the RPA can predict quite accurate band structure of bulk water and the hybridization effect\cite{cheng_calculation_2016}.
This is the case of $\text{HO}_{2}$ (Fig.~\ref{fig5}(c)).

The results will be compared with those calculated by DFT at several functionals (PBE0, HSE06 and BLYP).
It is noted that the band gap of bulk water is always underestimated at these functionals\cite{adriaanse_aqueous_2012,cheng_redox_2014,cheng_calculation_2016}, which will lead to a stronger hybridization between the redox couples and the band structure than in reality.
To account for this artificial hybridization in DFT, we use the VBM and CBM from these functionals as our model inputs.
We notice that, due to the larger band gap of bulk water in reality compared to that in DFT, the hybridization between $\text{A}^{-}$ and the band structure of bulk water in reality is expected to be smaller than that predicted by DFT. 
Therefore, we can safely utilize the experimental (or RPA’s) values of $-\text{IP}$ as the ``bare'' levels in our model.

The $\Delta$ is again a parameter, and we determine its value by aligning the $-\text{EA}$ level calculated by our model with that obtained from DFT (all the fitted values of $\Delta$ can be found in Table~\ref{tab3}).
This approach is analogous to what we did in last section, where the value of $\Delta$ is established by matching the level of the bound state calculated by our model with that by DFT.  

The calculated $-\text{IP}$ level of $\text{A}^{-}$, $-\text{EA}$ level of $\text{A}^{\bullet}$, and $U^{o}$ of the couple $\text{A}^{-}/\text{A}^{\bullet}$ by our model are shown in Fig.~\ref{fig5} (solid red lines for $-\text{IP}$, solid blue lines for $-\text{EA}$ and solid green lines for $U^{o}$).
The results obtained from DFT (dashed lines, the data is taken from Refs.~\onlinecite{cheng_redox_2014,cheng_calculation_2016}) are also included in the figure for the convenience of comparison.
The leftmost column in each subfigure shows the model inputs for $-\text{IP}$ and $\lambda$ (with $U^{o}=-\text{IP}+\lambda$ and $-\text{EA}=-\text{IP}+2\lambda$) for each case.

Fig.~\ref{fig5} shows that the hybridization significantly shifts the levels of the $-\text{IP}$, $-\text{EA}$, and $U^{o}$.
This can be seen by comparing with the levels in the leftmost column of each subfigure, which represent the model inputs.
We see that our model results again agree well with the DFT results, thus leading credence to our claim that the hybridization effect on electrochemical redox reactions can be effectively captured within the framework of the Haldane-Anderson model.

However, we also notice that the results for $\text{Cl}^{-}/\text{Cl}^{\bullet}$ and $\text{OH}^{-}/\text{OH}^{\bullet}$ hybridized with the band structure at BLYP deviate large from DFT.
In fact, such deviation is attributed to the failure of WBA in this case, which does not impact the above conclusion.
We discuss this in Appendix~\ref{apb}.
\section{Discussion}
\label{sec4}
\subsection{Implication to catalytic activity of semiconductor electrodes}
\label{sec4a}
The reaction $\text{OH}^{-}\rightarrow \text{OH}^{\bullet}+\text{e}^{-}$ occurring on the surface of $\text{TiO}_{2}$, shown in Sec.~\ref{sec3a}, is a model reaction for analyzing the possible origins of catalytic activities of semiconductor catalysts.

As evident from Fig.~\ref{fig4}(b), $\text{TiO}_{2}$ indeed facilitates the redox reaction $\text{OH}^{-}\rightarrow \text{OH}^{\bullet}+\text{e}^{-}$.
The increase of $\Delta$ can be interpreted as the ``turn on'' of the hybridization effect.
Thus, results in Fig.~\ref{fig4}(b) indicate that as the hybridization effect is ``turned on'', the resulting reorganization energy of the redox reaction decreases. 
According to the Marcus theory\cite{marcus_electrostatic_1956,marcus_theory_1965}, the activation energy of eletron transfer reactions is determined by both the free energy change and reorganization energy\cite{memming_semiconductor_2015}.
Consequently, a decrease in reorganization energy leads to a lower activation energy. 
Therefore, Fig.~\ref{fig4}(b) suggests that hybridization with semiconductor catalysts decreases the activation energy of the redox reaction.
This phenomenon has already been observed by DFT calculations\cite{cheng_identifying_2014}.

\begin{figure}[htb]
    \centering
    \includegraphics{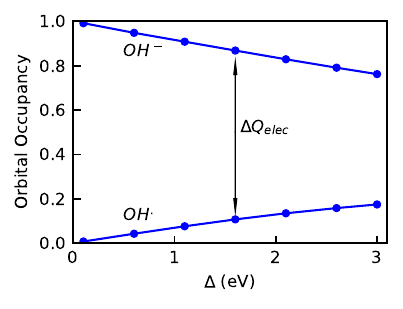}
    \caption{The occupancy of HOMO of $\text{OH}^{-}$ and LUMO of $\text{OH}^{\bullet}$, respectively, as a function of $\Delta$.}
    \label{fig6}
\end{figure}

To understand this phenomenon, we compute the occupancy of the HOMO of $\text{OH}^{-}$ and LUMO of $\text{OH}^{\bullet}$ after hybridization with $\text{TiO}_{2}$ (Fig.~\ref{fig6}).
Notably, the occupancy of HOMO of $\text{OH}^{-}$ deviates from 1 and decreases with an increase in $\Delta$.
Similarly, the occupancy of LUMO of $\text{OH}^{\bullet}$ becomes greater than 0 when we ``turn on'' the hybridization effect.
As a result, the oxidation of $\text{OH}^{-}$ to $\text{OH}^{\bullet}$ does not require the removal of a full electronic charge from $\text{OH}^{-}$ when the hybridization effect is ``turned on''.
Specifically, the charge that needs to be removed from $\text{OH}^{-}$ to complete the reaction is less than 1 and becomes significantly smaller as $\Delta$ increases (as shown by $\Delta Q_{\text{elec}}$ in Fig.~\ref{fig6}).
This reduction in transferred charge during redox reactions alleviates the response of the solvation shell, explaining why reorganization energy and the activation energy decrease as $\Delta$ is increased.

Thus, the origin of catalytic activities of semiconductor catalysts, within our framework, lies in the self-consistent treatment of the actual charge.
With this treatment, the actual amount of transferred charge during redox reactions diminishes as hybridization is stronger.
This reduction in transferred charge results in a smaller reorganization energy and lower activation energy, thereby facilitating the reaction.
Notably, this hybridization effect of semiconductor band structure on electrochemical redox reactions is very similar to that on charge state transitions in defect physics\cite{haldane_simple_1976,raebiger_charge_2008}.
In Appendix~\ref{apc}, by comparing with a control model, we further verify that the self-consistent treatment is indeed necessary to correctly describe the hybridization effect.
\subsection{Asymmetry in reorganization energies}
\label{sec4b}
When using Marcus theory to interpret the electrochemical experiments, one often uses a single reorganization energy for both oxidation and reduction processes.
However, it is reported that for some species, the reorganization energy should be different for oxidized and reduced species.
This asymmetry in reorganization energies is always ascribed to the different inner-sphere (i.e., intra-molecular) force constants for oxidized and reduced species\cite{hupp_driving-force_1984,tsirlina_asymmetry_1998,nazmutdinov_activation_1998,weaver_extant_2001,ignaczak_effects_2007,laborda_asymmetric_2012,laborda_asymmetric_2013,zeng_simple_2015,matyushov_q-model_2018}.
It is interesting that our results, as depicted in Fig.~\ref{fig4}(b), also reveal such asymmetry in reorganization energies (i.e., $\lambda_{O}^{\prime}<\lambda_{R}^{\prime}$, the prime here denotes that the reorganization energy is the computed one, to be distinguished from the one as model inputs), despite we employ a linear solvent response (Eq.~\ref{eq19}) and do not include the inner-sphere force constants in our model.
Here, we will use our model Hamiltonian to illustrate the alternative origin of the asymmetry in reorganization energies: the result of hybridization with materials with extended states.
We point out that the asymmetry induced by hybridization has already been observed in previous AIMD calculations by some of us\cite{adriaanse_aqueous_2012,cheng_redox_2014,liu_aqueous_2015,cheng_calculation_2016}.
In this work, we further identify the cause, as detailed below.

\begin{figure}[htb]
    \centering
    \includegraphics{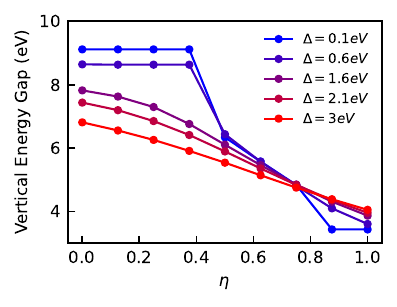}
    \caption{The vertical energy gap $\Delta E$ as a function of the coupling parameter $\eta$ for various values of $\Delta$.}
    \label{fig7}
\end{figure}

We first interpret the cause based on equations.
We obtain the free energy changes and then $\lambda^{\prime}$ by Eq.~\ref{eq33}.
So, we first check how the vertical energy gap $\Delta E$ changes as a function of $\eta$ (see Fig.~\ref{fig7}).
If $\Delta E$ exhibits a linear dependence on $\eta$, the free energy change obtained from TI integration can be computed by the mean of the values at the two endpoints (see Eq.~\ref{eq33}), resulting in $\lambda_{O}^{\prime}=\lambda_{R}^{\prime}$.
When $\Delta E$ is not linearly dependent on $\eta$, an asymmetry in $\lambda^{\prime}$ will be observed, which is the case shown in Fig.~\ref{fig7}, especially those with small $\Delta$.
Thus, elucidating the source of the nonlinear dependence of $\Delta E$ on $\eta$ is vital for understanding the asymmetry.
In fact, what we have constructed by the linear coupling method is a fictitious defect, whose interaction with the solvent is linearly coupled to that of $\text{OH}^{\bullet}$ and $\text{OH}^{-}$ (see Sec.~\ref{sec2d}).
This implies a linear dependence of the actual charge of this fictitious defect on $\eta$. 
This is evident from the first term in Eq.~\ref{eq34}.
However, this fictitious defect has to hybridize with the substrate's band structure, which is also evident from Eq.~\ref{eq34}.
As introduced in Sec.~\ref{sec2}, hybridization will result in a self-consistency requirement for the actual charge, and the consequence is the self-regulation of the charge.
So, when the change of the solvation shell is small (\textit{e.g.}, when $\eta$ changes from 0 to 0.2), self-regulation of the charge will have the tendency to keep itself to be constant (this is evident at the two end of $\eta$ as shown in Fig.~\ref{fig7}).
Thus, despite the initial linear dependence of the actual charge of the fictitious defect on $\eta$, hybridization disrupts this linearity due to the self-consistency requirement.
This is the reason why $\Delta E$ depends nonlinearly on $\eta$, and is the origin for the asymmetry in reorganization energies.

To delve deeper into this phenomenon, we now give a further discussion without the aid of equations.
Intuitively, to oxidize $\text{OH}^{-}$ to $\text{OH}^{\bullet}$, the change in formal charge of $-1\rightarrow 0$ means the actual charge of $\Delta Q\approx e^{-}$, where $e^{-}$ is the electronic charge, is needed to be transferred away.
This will lead to a reorganization energy of $\lambda^{\prime}\approx\lambda\cdot\Delta Q$.
Situation is similar when reducing $\text{OH}^{\bullet}$ back to $\text{OH}^{-}$.
The same charge $\Delta Q$ is needed, which will definitely result in a same reorganization energy of $\lambda^{\prime}$.
So, it seems that the reorganization energies will always be symmetric if we assume a linear response of the solvent.

Here, we argue that the key to understanding this phenomenon lies in the fact that the change of formal charge does not directly relate to the change of actual charge.
In fact, oxidation of $\text{OH}^{-}$ to $\text{OH}^{\bullet}$ means the transfer of the bound state charge, which can be approximated as $\sim Z(E_{\text{OH}^{-}}^{\text{eff}})$, depending on the effective defect level $E_{\text{OH}^{-}}^{\text{eff}}$ (see Sec.~\ref{sec2c}).
The remaining part of charge will respond with a self-regulation mechanism, after the bound state charge is transferred away.
Similarly, during the reduction of $\text{OH}^{\bullet}$ to $\text{OH}^{-}$, what needs to be transferred is also the bound state charge, but with a different value of $\sim Z(E_{\text{OH}^{\bullet}}^{\text{eff}})$.
The remaining part of charge will also self-respond after the bound state charge is transferred.
So, despite the total transferred charge $\Delta Q$ being identical for both oxidation and reduction processes, the key quantity\textemdash bound state charge $Z$\textemdash differs for oxidized state and reduced state.
By this point, we can at least say that the asymmetry in reorganization energies is not incomprehensible, since the charge transfer processes are indeed different for oxidation and reduction processes, once one recognize that the charge state transition isn't directly related to the change of actual charge.

\begin{figure}[htb]
    \centering
    \includegraphics{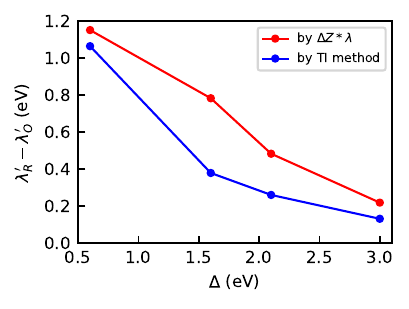}
    \caption{The difference of reorganization energies $\Delta \lambda^{\prime}=\lambda_{R}^{\prime}-\lambda_{O}^{\prime}$ computed by Eq.~\ref{eq36} (in red) and by Eq.~\ref{eq37} (in blue).}
    \label{fig8}
\end{figure}

To verify the above argument, we check how the solvent respond to the bound state charge.
We compute the extent of the difference in reorganization energies induced by the difference of bound state charges, by
\begin{equation}
    \Delta\lambda_{\text{bound}}^{\prime}=[Z(E_{\text{OH}^{\bullet}}^{\text{eff}})-Z(E_{\text{OH}^{-}}^{\text{eff}})]\lambda,
    \label{eq36}
\end{equation}
which is shown by the red line in Fig.~\ref{fig8} (the data is taken from the case of Sec.~\ref{sec3a}).
We also show the rigorous results obtained by TI method
\begin{equation}
    \Delta\lambda_{\text{rigorous}}^{\prime}=\lambda_{R,\text{TI}}^{\prime}-\lambda_{O,\text{TI}}^{\prime}
    \label{eq37}
\end{equation}
in blue line in the figure, which takes into account not only the bound state charge but also the charge in incoherent part ($\lambda_{R,\text{TI}}$ and $\lambda_{O,\text{TI}}$ are taken from Fig.~\ref{fig4}).
From Fig.~\ref{fig8}, we see that the results by Eq.~\ref{eq36} not only predicts the asymmetry of reorganization energies, but also gives an even larger $\Delta\lambda^{\prime}$: $\Delta\lambda_{\text{bound}}^{\prime}>\Delta\lambda_{\text{rigorous}}^{\prime}$.
This suggests that the different reorganization energies for oxidation and reduction processes are indeed caused by the different bound state charges needed for oxidation and reduction reactions.
And the reorganization energies caused by the charge in incoherent part may slightly ease the asymmetry between $\lambda_{O}^{\prime}$ and $\lambda_{R}^{\prime}$.

The above discussion demonstrate that, the asymmetry in reorganization energies induced by hybridization as observed by AIMD calculations\cite{adriaanse_aqueous_2012,cheng_redox_2014,liu_aqueous_2015,cheng_calculation_2016}, can be understood by a simple picture once one recognizes that the charge state transition cannot be interpreted as a change in actual charge, when the redox species interacts with materials with extended states.
\subsection{Effect of semiconductor band structure on redox reaction in strong coupling limit}
\label{sec4c}
Now, we want to qualitatively discuss the hybridization effect of semiconductor band structure on the redox reactions from a distinct perspective: the strong coupling limit.
Here, the strong coupling means the interaction between the species and the semiconductor substrate is so strong that the species becomes a part of the substrate.
We still take the reaction $\text{OH}^{-}\rightarrow\text{OH}^{\bullet}+\text{e}^{-}$ occurring on the surface of $\text{TiO}_{2}$ as an example.

\begin{figure}[htb]
    \centering
    \includegraphics[width=\linewidth]{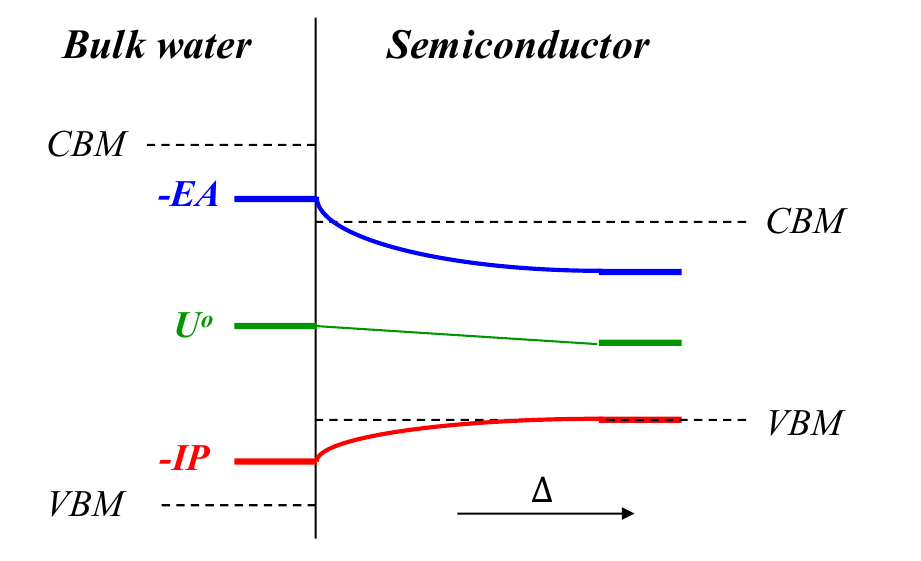}
    \caption{Schematic illustration of the effect of the semiconductor band structure on redox reaction in strong coupling limit.}
    \label{fig9}
\end{figure}

In this strong coupling limit, the $-\text{IP}$ level of $\text{OH}^{-}$ converges to the VBM of the semiconductor substrate, as it becomes an integral part of the substrate now.
For the $-\text{EA}$ level of $\text{OH}^{\bullet}$, it equals to the $-\text{EA}$ of lattice oxygen $\text{O}^{\bullet-}$, which is the $-\text{EA}$ of the substrate containing \textit{one hole}. 
The $-\text{EA}$ of the substrate containing one hole will be lower than the CBM of the substrate, owing to the enhanced electron attraction facilitated by the hole.
This means the $-\text{EA}$ level of $\text{OH}^{\bullet}$ will be lower than the CBM of the substrate.
We show this schematically in Fig.~\ref{fig9}.
For the redox potential $U^{o}$ of $\text{OH}^{-}/\text{OH}^{\bullet}$, we approximate it as the average of $-\text{IP}$ and $-\text{EA}$ level, since in this strong coupling case, it is a good approximation for the reorganization energy to be symmetric (see Fig.~\ref{fig4}, the larger $\Delta$ gives more symmetric $\lambda^{\prime}$).
Therefore, based on this discussion, the $-\text{IP}$, $-\text{EA}$ and $U^{o}$ levels exhibit a tendency to be confined by the VBM and CBM of the substrate.

It is interesting that we see the hybridization effect once again, albeit from a distinct perspective this time.
In this picture, the energy separating between $-\text{IP}$ and $-\text{EA}$ decreases due to the confinement of the band edges.
This phenomenon is consistent with our prior observations in Figs.~\ref{fig4} and ~\ref{fig5}.
In fact, the confinement effect of the semiconductor observed here is fundamentally the hybridization effect discussed in Sec.~\ref{sec4a}.
The reduction in the energy separation between $-\text{IP}$ and $-\text{EA}$ is attributed to the reduced transferred charge between $\text{A}^{-}$ and $\text{A}^{\bullet}$, which is the result of the hybridization effect.
\subsection{Comparison with related works in the literature}
\label{sec4d}
\subsubsection{Comparison with Hamiltonian models in electrochemistry}
\label{sec4d1}
There have been several works in electrocatalysis taking a model Hamiltonian approach to investigate the catalytic effect of electrodes, particularly for metallic electrodes.
Schmickler pioneered developing such a theoretical model for electrocatalysis\cite{schmickler_theory_1986}, where the chemisorption between aqueous species and metallic electrodes was described by the Newns-Anderson model\cite{newns_self-consistent_1969}.
In Schmickler's theory, chemisorption stabilizes the aqueous species, which can be described by the chemisorption function $\Delta (\epsilon)$, named by Newns\cite{newns_self-consistent_1969}.
This stabilization reduces the energy barrier for electron transfer reactions, and is regarded as a key factor to electrocatalysis\cite{huang_mixed_2020,santos_models_2022}.

For semiconductor electrodes, there are several related models.
Gao et al. used the tight-binding calculations for the solid and the extended-Hückel method for the coupling to the redox species at the interface\cite{gao_theory_2000,gao_theory_2000-1}.
Then, Davydov et al. used the Anderson Hamiltonian to describe the chemisorption on semiconductor surfaces in vacuum, and used the Green's function method to compute the energy level of adatom on the surface\cite{davydov_adsorption_2007,davydov_adsorption_2008}, which follows the way of the Haldane-Anderson model.
Schmickler et al. also used similar treatment to describe chemisorption in their eletrocatalytic model for semiconductor electrodes\cite{schmickler_adiabatic_2017}.
To describe the electron transfer reaction in their model, the solvent effect was accounted for through the reorganization energy $\lambda$ in response to the change of solute charge, following the Marcus theory.
With these treatments, the energy barrier of electron transfer reaction can be estimated, and the electronic hybridization effect on it can be understood\cite{schmickler_adiabatic_2017}.

Our model is consistent with Schmickler's model, but we focus on the equilibrium oxidized and reduced state.
Using the treatment introduced by Anderson\cite{anderson_model_1975}, a ``negative-\textit{U}'' term is obtained in the Hamiltonian.
With this term, the self-consistency effect can be explicitly studied.
We have shown that the self-consistency effect of the actual charge of the aqueous species is similar to the ``charge self-regulation'' mechanism in defect physics\cite{raebiger_charge_2008}.
And the self-consistent treatment indeed has interesting effects for semiconductor electrocatalysis.

Here, we further point out that, although it is intriguing to see whether this self-consistent treatment has similar effects for metallic electrodes, considering the tendency of delocalization in the electronic states of adsorbates on metallic electrodes\cite{fajin_need_2012,mom_modeling_2014}, the present framework of relating actual charge to formal charge may need to be modified.
\subsubsection{Comparison with catalysis theories in terms of adsorption energy}
\label{sec4d2}
In electrocatalysis, the catalytic effects of electrodes are usually understood in terms of adsorption energies of reaction intermediates\cite{hammer_why_1995,norskov_density_2011,xin_effects_2014,lee_prediction_2011,grimaud_double_2013,giordano_electronic_2022,suntivich_perovskite_2011,hwang_perovskites_2017,zhang_adsorption_2022,jiao_descriptors_2022,ooka_sabatier_2021,zhang_adsorption_2022}.
According to the Born-Haber cycle, the redox/protonation/hydrogenation energy of an elementary proton-coupled electron transfer (PCET) step can be obtained from adsorption energy differences between the reduced/protonated/hydrogenated and oxidized/deprotonated/dehydrogenated states\cite{cheng_redox_2009}.
Therefore, the redox potential that can be obtained from the present work is similar to the dehydrogenation energy that is obtained from adsorption energy differences. 

Combined with adsorption energies computed by DFT, dehydrogenation energies with reference to a gas-phase $\text{H}_{2}$ molecule can be computed using the computational hydrogen electrode (CHE) model\cite{norskov_origin_2004}.
In the CHE model, taking oxygen evolution reaction (OER) as an example, four successive steps of concerted proton and electron transfers are assumed, so that the reaction intermediates are considered to be $*\text{OH}$, $*\text{O}$, and $*\text{OOH}$, where $*$ stands for the binding site on the electrode surface.
With CHE model, the potential determining step can be identified, and the overpotential can be estimated\cite{norskov_origin_2004}.
It is further pointed out that there exists a scaling relationship between the adsorption energies of $*\text{OOH}$ and $*\text{OH}$, which will lead to a lowest possible theoretical overpotential for OER\cite{rossmeisl_electrolysis_2005,rossmeisl_electrolysis_2007,man_universality_2011,koper_thermodynamic_2011}.
It is noticed that the scaling relationship was justified also based on the assumption of the concerted proton and electron transfer\cite{man_universality_2011,koper_thermodynamic_2011}.

The above picture is prevalent in the field, and indeed provides useful insights into electrocatalysis.
However, it was reported that for many electrodes especially for oxides it is sequential PCET not concerted PCET that was observed\cite{koper_theory_2013-1,ooka_sabatier_2021,koper_theory_2013,chen_chemical_2013,gorlin_tracking_2017,yang_redefinition_2020,cheng_aligning_2014,warburton_theoretical_2022}.
Here, we emphasize that, by the framework presented in this work, we can study the reaction with sequential PCET mechanism and focus on redox reactions.
We have shown that an intuitively appealing picture of the catalytic effect of semiconductor electrodes on electron transfer can be obtained.
\section{Conclusions}
\label{sec5}
In summary, to understand the catalytic effect of semiconductor electrodes on redox reactions, we extend the Haldane-Anderson model to electrochemical systems by integrating the solvent effect inspired by the Holstein model.
With a simple treatment introduced by Anderson and a compatible framework proposed by Haldane and Anderson, the actual charge of the species in reduced and oxidized states can be treated in a self-consistent manner.
We verify that this self-consistent treatment is necessary to correctly describe the hybridization effect of band structure on redox reactions by comparing the model-calculated $-\text{IP}$, $-\text{EA}$, and redox potential with those obtained from DFT calculations. 
And we highlight that the self-consistency effect is the key to obtaining a fuller understanding on the origin of catalytic activities of semiconductor catalysts, and the origin of asymmetry in reorganization energies, as we demonstrate in the main text.

Given the complexity of electrochemical systems, there exist a few different perspectives in the literature concerning the effect of electrode's electronic structure on catalytic reactions. 
The parallel drawn in this work between electrochemical redox reactions and charge state transitions in defect physics suggests a promising avenue to comprehend band structure effects through the lens of defect physics.

While the model Hamiltonian approach may not yield first principles accuracy, it presents a clear picture compared to first principles calculation.
Notably, in situations where DFT cannot provide accurate results, the advantage of the model Hamiltonian method lies in its ability to offer us a clear physical picture for guidance.
Therefore, despite the increasing computational power available today, the significance of the model Hamiltonian method should not be overlooked.
\begin{acknowledgments}
J.C. acknowledges funding from the National Science Fund for Distinguished Young Scholars (Grant No.22225302), the National Natural Science Foundation of China (Grant Nos. 92161113, 21991151, 21991150, and 22021001), the Fundamental Research Funds for the Central Universities (Grant Nos. 20720220008, 20720220009, and 20720220010), Laboratory of AI for Electrochemistry (AI4EC), and IKKEM (Grant Nos. RD2023100101 and RD2022070501).
J.H. acknowledges the Initiative and Networking Fund of the Helmholtz Association (Grant No. VH-NG-1709) and European Research Council (ERC) Starting Grant (MESO-CAT, Grant agreement No. 101163405).
The authors thank Professor Hong Jiang at Peking University and Professor Xinguo Ren at Institute of Physics, Chinese Academy of Sciences, for helpful discussion.
\end{acknowledgments}
\section*{DATA AVAILABILITY}
The data that support the findings of this study are available from the corresponding author upon reasonable request.
\appendix
\section{Green's function method}
\label{apa}
For the single-particle effective Anderson Hamiltonian
\begin{equation}
    H=\sum\limits_{k}\epsilon_{k}n_{k}+\sum\limits_{m}E_{m}n_{m}+\sum\limits_{k,m}V_{km}c_{k}^{\dagger}c_{m}+V_{mk}c_{m}^{\dagger}c_{k},
\end{equation}
the Green's function is
\begin{equation}
    (\omega -H)G=I,
\end{equation}
where $\omega$ is variable as used in Sec.~\ref{sec2a}, $I$ is the unity.
Use the eigenstates of system before hybridization as basis, the matrix equation can be written as:
\begin{equation}
    (\omega -E_{m})G_{mm}-\sum\limits_{k}V_{mk}G_{km}=1,
    \label{eqa3}
\end{equation}
\begin{equation}
    (\omega -\epsilon_{k})G_{km}-V_{km}G_{mm}=0,
    \label{eqa4}
\end{equation}
\begin{equation}
    (\omega-E_{m})G_{mk}-\sum\limits_{k^{'}}V_{mk^{'}}G_{k^{'}k}=0,
\end{equation}
\begin{equation}
    (\omega-\epsilon_{k^{'}})G_{k^{'}k}-V_{k^{'}m}G_{mk}=\delta_{k^{'}k}.
\end{equation}
From Eqs.~\ref{eqa3} and~\ref{eqa4}, we can solve for $G_{mm}$:
\begin{eqnarray}
    G_{mm}=\frac{1}{\omega-E_{m}-\sum\limits_{k}\frac{\lvert V_{mk}\rvert^{2}}{\omega-\epsilon_{k}}},
    \label{eqa7}
\end{eqnarray}
from which we define the self-energy $\Sigma (\omega)$:
\begin{equation}
    \Sigma (\omega)=\sum\limits_{k}\frac{\lvert V_{mk} \rvert ^{2}}{\omega-\epsilon_{k}}.
    \label{eqa8}
\end{equation}
Eqs.~\ref{eqa7} and~\ref{eqa8} are exactly Eqs.~\ref{eq5} and~\ref{eq6} in the main text.
The self-energy can be further written as
\begin{equation}
    \Sigma (\omega)=\lim_{s \to 0} \sum\limits_{k}\frac{\lvert V_{mk}\rvert ^{2}}{\omega-\epsilon_{k}+is}=\lim_{s \to 0} \sum\limits_{k}\frac{\lvert V_{mk}\rvert ^{2}[(\omega-\epsilon_{k})-is]}{(\omega-\epsilon_{k})^{2}+s^{2}},
    \label{eqa9}
\end{equation}
which contains both real and imaginary parts.
We define
\begin{equation}
    \Sigma^{'}(\omega)=\lim_{s \to 0} \sum\limits_{k}\frac{\lvert V_{mk}\rvert ^{2}(\omega-\epsilon_{k})}{(\omega-\epsilon_{k})^{2}+s^{2}},
    \label{eqa10}
\end{equation}
\begin{equation}
    \Delta(\omega)=\lim_{s \to 0} \sum\limits_{k}\frac{\lvert V_{mk}\rvert ^{2}s}{(\omega-\epsilon_{k})^{2}+s^{2}}.
    \label{eqa11}
\end{equation}
From Eqs.~\ref{eqa9} to ~\ref{eqa11} we obtain Eq.~\ref{eq7} in the main text.

To find out the DOS localized in the defect site, here we define $\ket{n}$ to be the eigenstates of the system after hybridization, and $\epsilon_{n}$ to be the corresponding eigenvalues.
Then, the DOS in the defect site is
\begin{equation}
    \rho_{m}(\omega)=\sum\limits_{n}\lvert\langle m\vert n\rangle\rvert^{2}\delta (\epsilon_{n}-\omega).
\end{equation}
Using the Lorentzian function to approximate the $\delta$ function, we have:
\begin{equation}
    \rho_{m}(\omega)=\frac{1}{\pi}\lim_{s \to 0}\sum\limits_{n}\lvert\langle m\vert n\rangle\rvert^{2}\frac{s}{(\omega-\epsilon_{n})^{2}+s^{2}}.
    \label{eqa13}
\end{equation}
Before linking the DOS to the Green's function, we need some further modification for $G_{mm}$.
We notice that
\begin{equation}
    G_{mm}=\langle m\lvert G\rvert m\rangle=\sum\limits_{n}\langle m\lvert n\rangle \langle n\lvert G \rvert n\rangle\langle n\rvert m\rangle,
    \label{eqa14}
\end{equation}
where we use the completeness relation of the basis $\ket{n}$:
\begin{equation}
    \sum\limits_{n}\lvert n\rangle \langle n\rvert=I.
\end{equation}
In the representation of the eigenstates $\ket{n}$ of the system after hybridization, $G$ is diagonal:
\begin{equation}
    G_{nn}=\lim_{s \to 0}\frac{1}{\omega-\epsilon_{n}+is}=\lim_{s \to 0}\frac{\omega-\epsilon_{n}-is}{(\omega-\epsilon_{n})^{2}+s^{2}}.
    \label{eqa16}
\end{equation}
Substituting Eq.~\ref{eqa16} into Eq.~\ref{eqa14}, $G_{mm}$ can be rewritten as:
\begin{equation}
    G_{mm}=\lim_{s \to 0}\sum\limits_{n}\lvert\langle m\vert n\rangle\rvert^{2}\frac{\omega-\epsilon_{n}-is}{(\omega-\epsilon_{n})^{2}+s^{2}}.
    \label{eqa17}
\end{equation}
Now, the relation between the DOS and the Green's function is clear.
Comparing Eqs.~\ref{eqa13} and~\ref{eqa17}, we found the famous equation
\begin{equation}
    \rho_{m}(\omega)=-\frac{1}{\pi}\text{Im}G_{mm}.
\end{equation}
Therefore, we can use the Green's function to calculate the occupancy and the energy level of the orbital by
\begin{equation}
    \int_{-\infty}^{\epsilon_{F}}-\frac{1}{\pi}\text{Im}G_{mm}(\omega)\,d\omega
\end{equation}
and
\begin{equation}
    \int_{-\infty}^{\epsilon_{F}}-\frac{\omega}{\pi}\text{Im}G_{mm}(\omega)\,d\omega,
\end{equation}
respectively.
\section{Origin of the deviation from results of BLYP}
\label{apb}
In Fig.~\ref{fig5} we observe that results for $\text{Cl}^{-}/\text{Cl}^{\bullet}$ and $\text{OH}^{-}/\text{OH}^{\bullet}$ hybridized with the band structure of BLYP deviate large from DFT.
We see that the VBM of BLYP is much higher than that of PBE0 and HSE06, and push the $-\text{IP}$ level much higher than that of PBE0 and HSE06.
However, our model cannot push the $-\text{IP}$ level effectively, when using the electronic coupling $\Delta$ describing the shift of $-\text{EA}$ level suitably.
This suggests that it is no longer a good approximation that the electronic coupling $\Delta$ between $-\text{IP}$ level and valence band is equal to that between $-\text{EA}$ level and conduction band.
That is, the WBA is no longer suitable for the case of BLYP.
In fact, the DOS of bulk water presented in Ref.~\onlinecite{zhang_modeling_2021} suggests that for bulk water, its DOS of the valence band is much larger than that of the conduction band.
Although the WBA is a good approximation for PBE0 and HSE06, Fig.~\ref{fig5} suggests that it is no longer the case for BLYP.

\begin{figure*}[htb]
    \centering
    \includegraphics{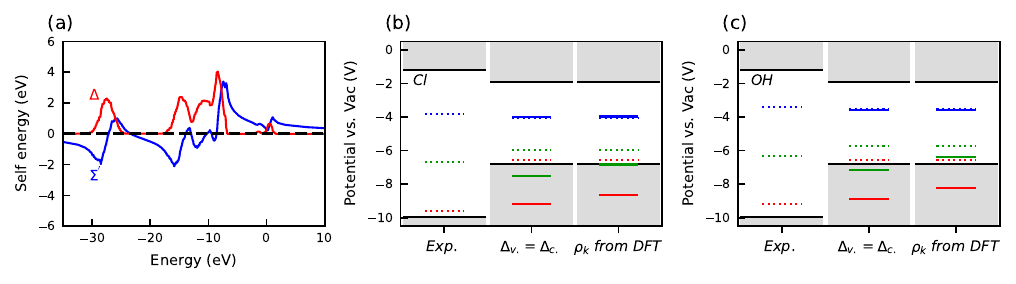}
    \caption{(a) The self-energy $\Sigma(\omega)$ with DOS from Ref.~\onlinecite{zhang_modeling_2021}. 
    (b) and (c) Level diagram of redox ion A interacts with the band structure of bulk water with the electronic coupling $\Delta$ taken as constant or with the DOS taken from Ref.~\onlinecite{zhang_modeling_2021}, as denoted in the figure.
    A represents Cl for (b), and represents OH for (c).
    The solid red, green, and blue lines corresponds to the $-\text{IP}$ level of $\text{A}^{-}$, $U^{o}$ of the couple $\text{A}^{-}/\text{A}^{\bullet}$, and $-\text{EA}$ level of $\text{A}^{\bullet}$.
    The dashed lines correspond to the values taken from DFT (the data is taken from Refs.\onlinecite{cheng_calculation_2016,cheng_redox_2014}).}
    \label{figs1}
\end{figure*}

To see whether the deviation observed in Fig.~\ref{fig5} is attributed to WBA, we utilize the DOS from Ref.~\onlinecite{zhang_modeling_2021} to replace the constant $\rho_{k}$ used in Eq.~\ref{eq8}.
Here, the $k$, $m$ independent $V_{km}$ is still employed.
The resulting self-energy $\Sigma(\omega)$ is shown in Fig.~\ref{figs1}(a).

With this $\Sigma(\omega)$, the calculated $-\text{IP}$ (solid red lines), $-\text{EA}$ (solid blue lines), and $U^{o}$ (solid green lines) are shown in the rightmost column of Figs.~\ref{figs1}(b) and (c).
We also show the results obtained by WBA in Fig.~\ref{figs1} for the convenience of comparison.
We can see that, by replacing the WBA with the DOS from Ref.~\onlinecite{zhang_modeling_2021}, the $U^{o}$ and $-\text{IP}$ are effectively uplifted for both the cases of $\text{Cl}$ and $\text{OH}$, reducing the deviations from the results of BLYP.
This suggests that the WBA indeed is not a suitable approximation in this case.
However, we also notice that the deviations from BLYP are still large.
Here, we attribute this to the failure of the approximation that the $V_{km}$ is independent on $k$, $m$.
In reality, the negatively charged anions of $\text{OH}^{-}$ and $\text{Cl}^{-}$ attract the positively charged $\text{H}^{+}$ in $\text{H}_{2}\text{O}$, while repelling the negatively charged $\text{O}^{2-}$ in $\text{H}_{2}\text{O}$.
This results in the anions being closer to $\text{H}^{+}$ than to $\text{O}^{2-}$ in water.
Consequently, $\text{Cl}^{-}$ and $\text{OH}^{-}$ hybridize more effectively with the orbitals of $\text{H}^{+}$ than with those of $\text{O}^{2-}$, and the $V_{km}$ is definitely depend on $k$ and $m$.
However, this level of complexity is beyond the scope of our model Hamiltonian here, and would require $ab\ initio$ calculations to capture it accurately.
\section{Necessity of self-consistent treatment of actual charge}
\label{apc}
In Sec.~\ref{sec4a}, we argue that the key to capturing the hybridization effect is the self-consistent treatment of actual charge.
To validate this, we develop a control model in which the occupancy of the orbital $\ket{m\sigma}$ of $\text{A}^{-}$ is fixed to be 1, while that of $\text{A}^{\bullet}$ is fixed to be 0:
\begin{equation}
    \langle n_{m\sigma} \rangle_{\text{A}^{-}}=1,
\end{equation}
\begin{equation}
    \langle n_{m\sigma} \rangle_{\text{A}^{\bullet}}=0.
\end{equation}
With these assignments, the Hamiltonian can be directly solved using Green's function method without the need for self-consistent calculations.

\begin{figure*}[htb]
    \centering
    \includegraphics{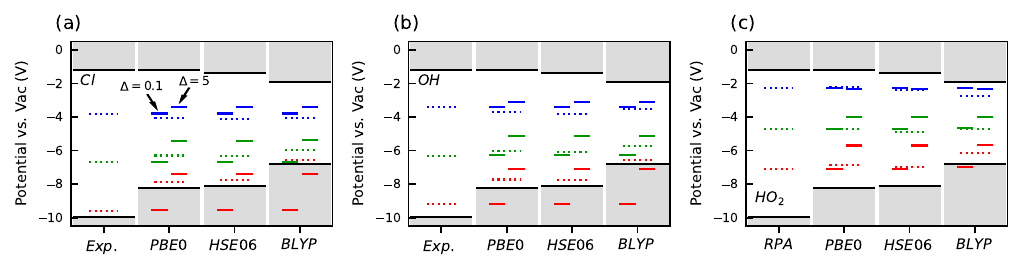}
    \caption{Level diagram of redox ion A interacts with the band structure of bulk water at different functionals, with A denotes (a) Cl, (b) OH and (c) $\text{HO}_{2}$.
    The solid red, green, and blue lines correspond to the $-\text{IP}$ level of $\text{A}^{-}$, $U^{o}$ of the couple $\text{A}^{-}/\text{A}^{\bullet}$, and $-\text{EA}$ level of $\text{A}^{\bullet}$ calculated by the control model, respectively, with electronic coupling $\Delta$ taken as 0.1 and 5eV, see main text.
    The dashed lines correspond to the values taken from DFT (the data is taken from Refs.\onlinecite{cheng_calculation_2016,cheng_redox_2014}).
    The band structure of bulk water are calculated by PBE0, HSE06 and BLYP as denoted in the figure, with corresponding VBM and CBM plotted by black solid lines.}
    \label{figs2}
\end{figure*}

We use the control model to calculate the couples $\text{Cl}^{-}/\text{Cl}^{\bullet}$, $\text{OH}^{-}/\text{OH}^{\bullet}$, and $\text{HO}_{2}^{-}/\text{HO}_{2}^{\bullet}$ hybridized with the band structure of bulk water.
The results computed by two distinct values of $\Delta$ ($\Delta=0.1$ eV and $\Delta=5$ eV) are shown in Fig.~\ref{figs2}.
We observe that when $\Delta=0.1$ eV, all the $-\text{IP}$, $-\text{EA}$ and $U^{o}$ remain almost unchanged relative to the values before hybridization.
When increasing $\Delta$ to 5 eV, all the $-\text{IP}$, $-\text{EA}$ and $U^{o}$ shift significantly.
It is interesting to notice that, for the case of Cl and OH, increasing $\Delta$ from 0.1 to 5 eV will uplift all the $-\text{IP}$, $-\text{EA}$, and $U^{o}$, leading to a more pronounced deviation of $-\text{EA}$ from the DFT results.
Consequently, in these cases, the $-\text{EA}$ predicted by the control model can no way match the DFT results.
This is quite different from the results shown in the main text (Fig.~\ref{fig5}), and suggesting the failure of the control model in describing the hybridization effect.
So, we conclude from these results that treating the actual charge and the solvent reorganization in a self-consistent way, as we do in the main text, is pivotal in correctly describing the hybridization effect.
\bibliography{1_ref_gujian_paper1}
\end{document}